\documentclass[pdflatex,sn-nature,oneside]{sn-jnl}
\usepackage{graphicx}%
\usepackage{multirow}%
\usepackage{amsmath,amssymb,amsfonts}%
\usepackage{amsthm}%
\usepackage{caption}
\usepackage{manyfoot}%
\usepackage{braket}
\usepackage{mathrsfs}%
\usepackage[title]{appendix}%
\usepackage{xcolor}%
\usepackage{textcomp}%
\usepackage{booktabs}%
\usepackage{algorithm}%
\usepackage{algorithmicx}%
\usepackage{algpseudocode}%
\usepackage{listings}%

\renewcommand{\thefigure}{\arabic{figure}}
\DeclareCaptionFont{snfig}{\fontsize{8}{9.5}\selectfont}

\makeatletter
\@twosidefalse
\makeatother

\usepackage[mathlines]{lineno} 

\usepackage{etoolbox}                

\usepackage{setspace}
\usepackage{geometry}
\AtBeginDocument{%
  \newgeometry{
    a4paper,
    left=30mm,
    right=30mm,
    top=25mm,
    bottom=25mm,
    includeheadfoot
  }%
  \setstretch{1.15}
}

\begin{document}
\captionsetup[figure]{labelfont={bf},labelformat={default},labelsep=period,name={Figure}}

\title{Strong-field ionization of atoms with bright squeezed vacuum light}

\author[1]{Haodong Liu}
\author[1]{Xiaoxiao Long}
\author[1]{Peizeng Li}
\author[1]{Zijian Lyu}
\author[1,2,3,*]{Yunquan Liu}

\affil[1]{State Key Laboratory for Mesoscopic Physics and Frontiers Science Center for Nano-optoelectronics, School of Physics, Peking University, Beijing, 100871, China}

\affil[2]{Collaborative Innovation Center of Extreme Optics, Shanxi University, Taiyuan, 030006, China}
\affil[3]{Peking University Yangtze Delta Institute of Optoelectronics, Nantong, Jiangsu, 226010, China}

\affil[*]{Yunquan.Liu@pku.edu.cn}

\flushbottom

\abstract
{Strong-field ionization is the cornerstone of attosecond physics, which has been extensively studied under coherent-state driving. Recently, the interface between attosecond physics and quantum optics has emerged as a new frontier. Yet, owing to experimental limitations, the role of the quantum nature of light in atomic strong-field ionization has remained unexplored. Here, we demonstrate strong-field ionization of xenon atoms driven by bright squeezed vacuum (BSV) with average pulse energy up to 10\,\textmu J. We show that, as a nonclassical state with zero mean field and strong intensity fluctuations, BSV selectively enhances the spider-like holographic structures in the photoelectron momentum distributions. Using a quantum-light-corrected quantum-trajectory Monte Carlo (q-QTMC) model, we attribute this effect to the intrinsic coherence of trajectory pairs emitted within the same subcycle field fluctuation. These dynamically correlated paths exhibit enhanced phase stability and remain robust against dephasing, whereas asynchronous paths are filtered out by field noise. Our results reveal a quantum-fluctuation-induced mechanism for coherence protection in strong-field processes, positioning BSV as an effective coherence filter and establishing a new regime of quantum-enabled noise-resilient ultrafast dynamics.

}

\maketitle
\newpage

\linenumbers%
\modulolinenumbers[10]
\section*{Main}

Strong-field physics investigates the interaction between intense laser pulses and matter. Among its fundamental processes is strong-field ionization of atoms. The basic picture of strong field ionization is understood within the pioneering work by Keldysh theory\cite{1965-keldysh-ionizationfieldstrong}. In the Keldysh framework, photoionization can be classified as multiphoton ionization or tunneling ionization, depending on the value of the Keldysh parameter $\gamma$. Specifically, multiphoton ionization dominates when $\gamma < 1$ while tunneling ionization takes place when $\gamma > 1$. In the intermediate regime, above threshold ionization (ATI) would take place, in which electrons can absorb more photons than the minimum required to ionize an atom \cite{1979-agostini-freefreetransitionsfollowing}. Strong-field ionization has laid the foundation for ultrafast science, enabling the generation of attosecond pulses through high-order harmonic generation (HHG) \cite{1988-ferray-multipleharmonicconversion1064,1994-lewenstein-theoryhighharmonicgeneration,1993-corkum-plasmaperspectivestrong} and facilitating the study of electron dynamics in atoms and molecules\cite{2004-itatani-tomographicimagingmolecular,2011-huismans-timeresolvedholographyphotoelectrons}. In the strong-field ionization process, the photoelectron momentum distribution (PMD) encodes rich electron dynamics. Depending on whether the electrons are released within the same optical cycle or across different cycles, the photoelectron patterns are classified as intracycle or intercycle interferences\cite{2006-arbo-timedoubleslitinterferences}. Intercycle interference produces the well-known concentric ATI rings. The intracycle interference gives rise to a set of holographic structures, including those arising from forward and backward electron rescattering\cite{2011-bian-subcycleinterferencedynamics,2012-bian-attosecondtimeresolvedimaging,2014-li-recollisioninducedsubcycleinterference}. Among them, the forward-scattered interference produces a characteristic spider-like pattern, which underpins photoelectron holography \cite{2004-spanner-readingdiffractionimages,2011-huismans-timeresolvedholographyphotoelectrons}, which is a powerful approach for imaging molecular structure and capturing ultrafast electron dynamics\cite{2014-meckel-signaturescontinuumelectron,2018-porat-attosecondtimeresolvedphotoelectron}. It arises from the interference between a reference electron wave, which reaches the detector without scattering, and a signal wave that undergoes scattering by the parent ion during strong-field ionization process\cite{2011-huismans-timeresolvedholographyphotoelectrons,2012-hickstein-directvisualizationlaserdriven,2024-khurelbaatar-strongfieldphotoelectronholography}. Because the holographic fringes are governed by the electron de Broglie wavelength and the sub-cycle emission timing, decoding the hologram embedded in the PMD allows to retrieve the structural and temporal information with sub-ångström spatial and attosecond temporal resolution.

The current understanding of strong-field processes relies on the assumption that the driving field is classical coherent-state light \cite{1993-corkum-plasmaperspectivestrong,1994-walker-precisionmeasurementstrong,2001-salieres-feynmanspathintegralapproach}. However, recent developments have begun to extend this classical-field paradigm by incorporating the quantum nature of light\cite{2020-gorlach-quantumopticalnaturehigh,2022-rivera-dean-lightmatterentanglementabovethreshold,2023-bhattacharya-stronglaserfield,2023-eventzur-photonstatisticsforceultrafast,2023-gorlach-highharmonicgenerationdriven,2024-lange-electroncorrelationinducednonclassicalitylight,2024-rivera-dean-squeezedstateslight,2024-stammer-absencequantumoptical}. The crossover between quantum optics with attosecond physics has been the new frontiers. The quantum light, such as bright squeezed vacuum (BSV) with zero mean field and strong fluctuations\cite{1997-scully-quantumoptics,2001-wolfgang-quantumopticsphase}, gives rise to new behaviors. For example, the HHG driven by BSV has revealed novel scaling laws, indicating that large intensity fluctuations can significantly enhance non-perturbative processes\cite{2023-gorlach-highharmonicgenerationdriven}. In two-color configurations combining the BSV and coherent-state light, the photon-bunched sidebands have been observed, suggesting a viable pathway for generating quantum light sources in the extreme ultraviolet (XUV) regime\cite{2025-lemieux-photonbunchinghighharmonic,2025-tzur-attosecondresolvedquantumfluctuations}. Importantly, the quantum statistical properties of the driving light is expected to be transferred to the photoelectrons, offering a direct pathway to generate quantum electron beams with tailored statistics \cite{2025-lyu-effectphotonquantum}. This concept has been experimentally demonstrated, for instance, for multiphoton ionization of metal tips, where the quantum statistical properties of BSV can be imprinted onto photoelectrons, opening a potential path toward engineered quantum electron sources\cite{2024-heimerl-multiphotonelectronemission,2025-heimerl-quantumlightdrives}. While quantum-light-driven strong-field physics draws much attention, the experimental study on strong-field ionization of atoms has not been conducted. According to our best knowledge, the intensity of reported BSV generation is not enough to ionize the atoms\cite{2012-sh.iskhakov-superbunchedbrightsqueezed,2014-perez-brightsqueezedvacuumsource}. Little is known about how the quantum nature of BSV affects the phase coherence of photoelectrons, which is a critical factor in the interferometric features. In particular, as is known, BSV has pronounced intensity fluctuations. This raises several crucial questions, e.g., does the interference structure survive in the noise of a quantum field? How does the quantum nature of the light alter the subcycle emission dynamics and trajectory correlations?

To address those fundamental questions, we experimentally investigate strong-field ionization of xenon atoms driven by intense BSV and analyze the resulting photoelectron holograms. Since the formation of holographic fringes relies sensitively on the coherence between the tunneling electron pathways, the interference structure would serve as a direct probe of how quantum-state light modifies the electron dynamics. In this way, photoelectron holography offers a unique route into the interplay between quantum light and ultrafast electron dynamics.

\begin{figure}[t]
    \centering
    \includegraphics[width=\textwidth]{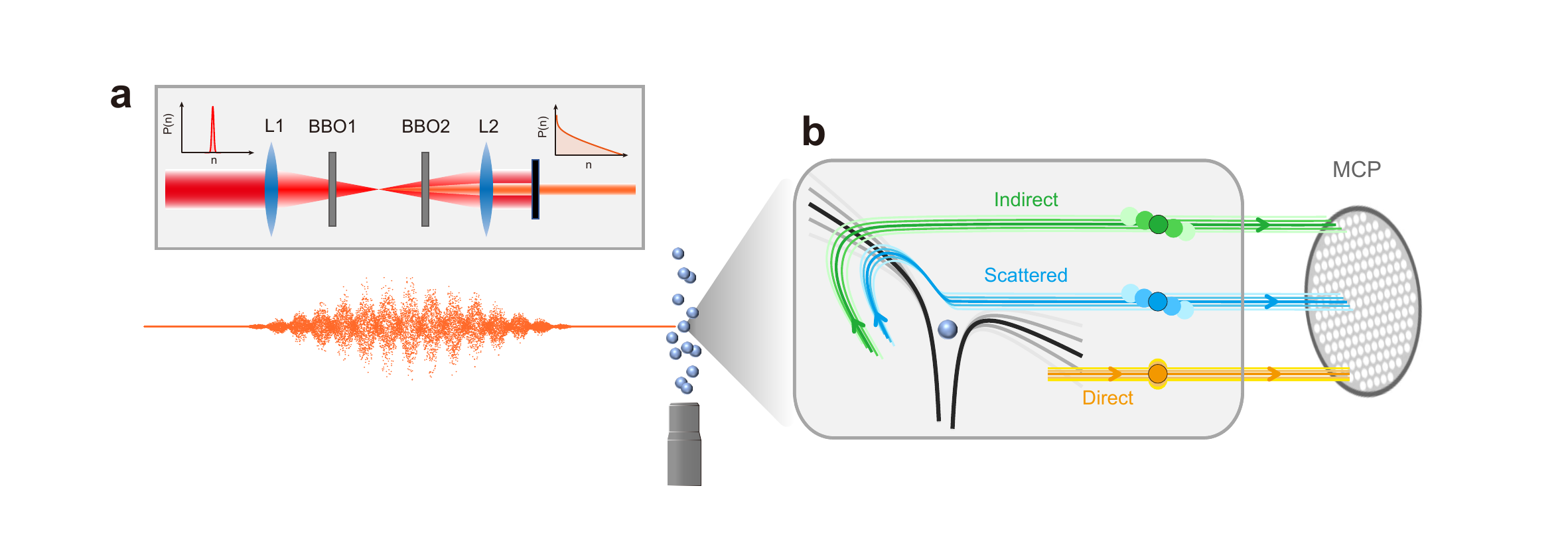}
    \caption{\textbf{Photoelectron holography in strong field ionization with bright squeezed vacuum.} \textbf{a}, Experimental setup for generating bright squeezed vacuum (BSV). An $800\,\mathrm{nm}$ laser beam (Poissonian photon statistics) is focused between two symmetrically aligned BBO crystals via a 4\textit{f} system. BSV (super-Poissonian photon statistics) centered at $1600\,\mathrm{nm}$ is generated in the BBO crystals. \textbf{b}, The generated BSV is focused onto a supersonic atomic beam, where electrons from the atoms undergo strong-field ionization via multiple quantum pathways. Each path accumulates a different phase due to the atomic Coulomb potential and results in distinct contributions to PMD. }
    \label{fig. 1}
\end{figure}

Experimentally, the intense BSV light is generated through the degenerate spontaneous parametric down-conversion (SPDC) with double-crystal geometry (see experimental method for the details), as illustrated in Fig. \ref{fig. 1}a. We have obtained near-single-mode BSV with average pulse energies of up to $10\,\mathrm{\mu J}$, which is sufficiently intense to induce strong-field ionization in xenon atoms. To probe BSV-driven photoelectron holography, we perform strong-field ionization experiments using a Cold Target Recoil Ion Momentum Spectroscopy (COLTRIMS) setup\cite{2003-ullrich-recoilionelectronmomentum,2000-dorner-coldtargetrecoil}. In Fig. \ref{fig. 1}b, we illustrate the microscopic subcycle electron dynamics under BSV driving light. In a cycle of BSV light, most of the released electrons can be categorized as direct, scattered, or indirect trajectories depending on the ionization instant\cite{2012-bian-attosecondtimeresolvedimaging,2011-bian-subcycleinterferencedynamics}. As is known, the BSV light follows super-Poissonian photon statistics, leading to strong shot-to-shot fluctuations of instantaneous intensity. Consequently, the field-suppressed Coulomb potential and those photoionized electron trajectories fluctuate with the quantum light field accordingly. The photoelectron momentum distributions are recorded with two-dimensional position-sensitive detectors (see experimental method for details), which allows us to investigate how the quantum nature of BSV is imprinted on photoelectron interference. 

\begin{figure}[t]
    \centering
    \includegraphics[width=\textwidth]{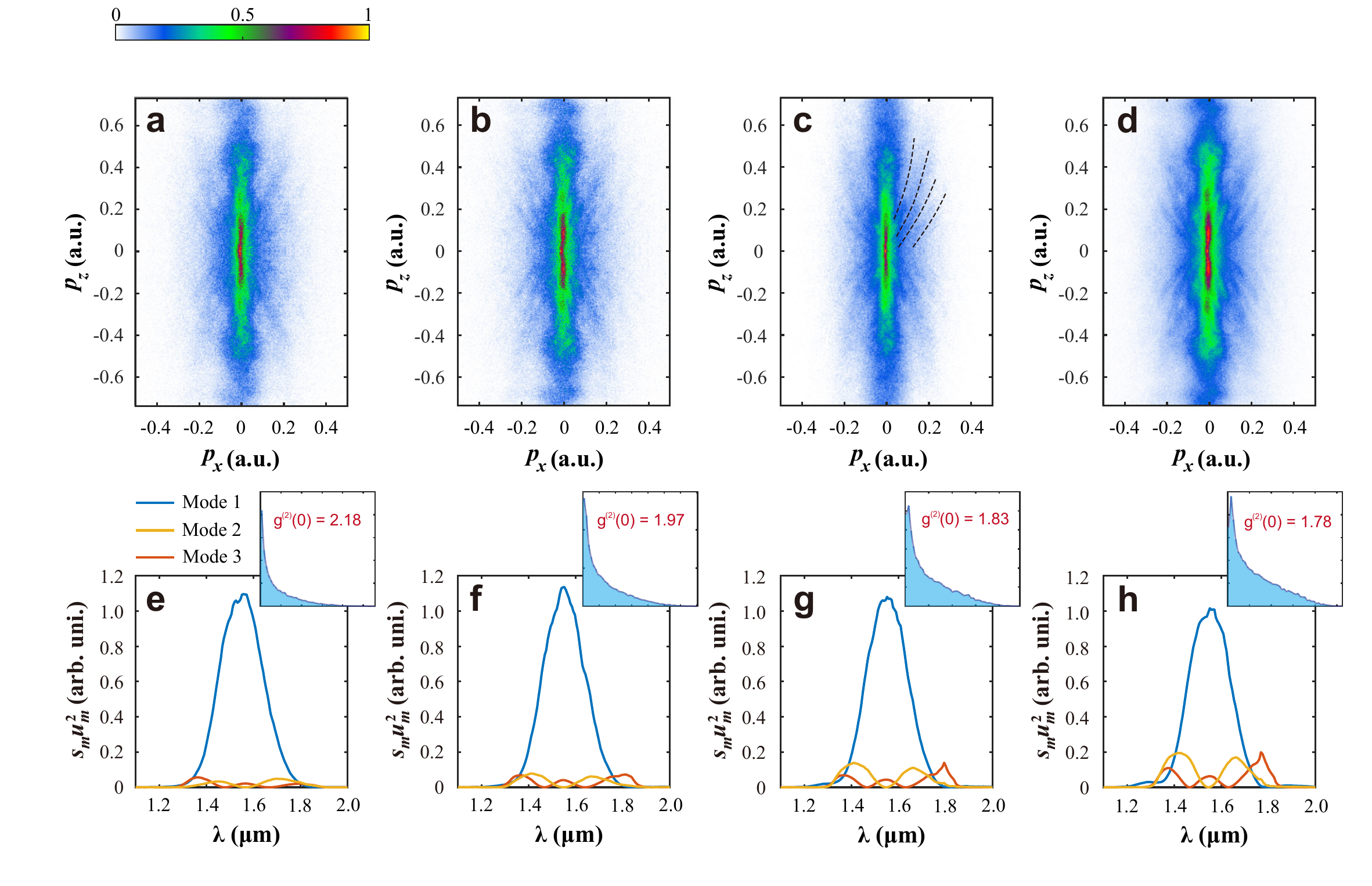}
    \caption{\textbf{Measured PMDs driven by BSV pulses with respect to the quantum statistics.} Panels \textbf{a-d}, show the results for the averaged pulse energies of $3.3\,\mathrm{\mu J}$, $5\,\mathrm{\mu J}$, $8.3\,\mathrm{\mu J}$, and $10\,\mathrm{\mu J}$, respectively. The corresponding average peak intensities are $0.53\times 10^{14}\,\mathrm{W/cm^2}$, $0.80\times 10^{14}\,\mathrm{W/cm^2}$, $1.33\times 10^{14}\,\mathrm{W/cm^2}$, and $1.60\times 10^{14}\,\mathrm{W/cm^2}$. Dashed lines in (\textbf{c}) indicate the spider-like structures. \textbf{e-h}, The first three contributing spectral modes of the BSV source are shown via modal decomposition. Each panel corresponds to the BSV driving field used in the photoelectron measurement shown directly above. In \textbf{e-h}, the insets display the measured photon-number distributions corresponding to each BSV pulse, from which the second-order correlation function $g^{(2)}$ is extracted.}
    \label{fig. 2}
\end{figure}

In Figures \ref{fig. 2}a-d, we present the measured PMDs using BSV pulses with average pulse energies of $3.3$, $5.0$, $8.3$, and $10\,\mathrm{\mu J}$, respectively. Notably, all spectra show the suppression of ATI peaks and reveal a ring-like inter-cycle interference structure that is typically prominent under the coherent-state driving\cite{2011-huismans-timeresolvedholographyphotoelectrons}. This suppression of ATI can be easily understood with the presence of strong pulse-to-pulse field fluctuations in BSV, which effectively disrupt the intercycle phase coherence necessary for ATI peak formation\cite{2023-fang-strongfieldionizationhydrogen}. Complementarily, shot-to-shot variations induced by BSV intensity fluctuations can also radially smear out the ATI rings by blurring the energy positions because of the ac-Stark shift.

Interestingly, the spider-like interference structure robustly persists across all measurements, as highlighted by dashed lines in Fig. \ref{fig. 2}c. Such structures are associated with photoelectron holography driven by the coherent-state light\cite{2011-huismans-timeresolvedholographyphotoelectrons,2011-bian-subcycleinterferencedynamics}, which is attributed to the interference between indirect and forward-scattered (hereafter referred to as scattered) electron trajectories. The distinct propagation dynamics of these trajectories in the combined laser and Coulomb fields are illustrated schematically in Fig. \ref{fig. 1}b. The persistence of this feature under strong quantum fluctuations suggests the presence of inherently coherence-resilient interference pathways, providing a concrete experimental pathway toward the quantum-noise-resilient imaging modalities.

We have also performed single-shot spectral characterization of BSV pulses. From the measured spectra, we can perform the modal decomposition that yields the spectral shape $u_m$ and corresponding weight $s_m$ of each mode (as described in Methods). In Figures \ref{fig. 2}e-h, we present the first three dominant spectral modes—modes 1, 2, and 3— for every experimental condition. Among them, mode 1 contributes most significantly. Since strong-field ionization is a highly nonlinear process, the PMDs are predominantly determined by this first mode. In parallel, the photon-number distributions of BSV pulses are measured. We then obtained the corresponding second-order correlation function $g^{(2)}$, as illustrated in the upper-right corner in Figures \ref{fig. 2}e-h. The corresponding values of $g^{(2)}$ are $2.18$, $1.97$, $1.83$, and $1.78$, respectively. The $g^{(2)}$ decreases with increasing pulse energy of BSV.

\begin{figure}[htbp]
    \centering
    \includegraphics[width=\textwidth]{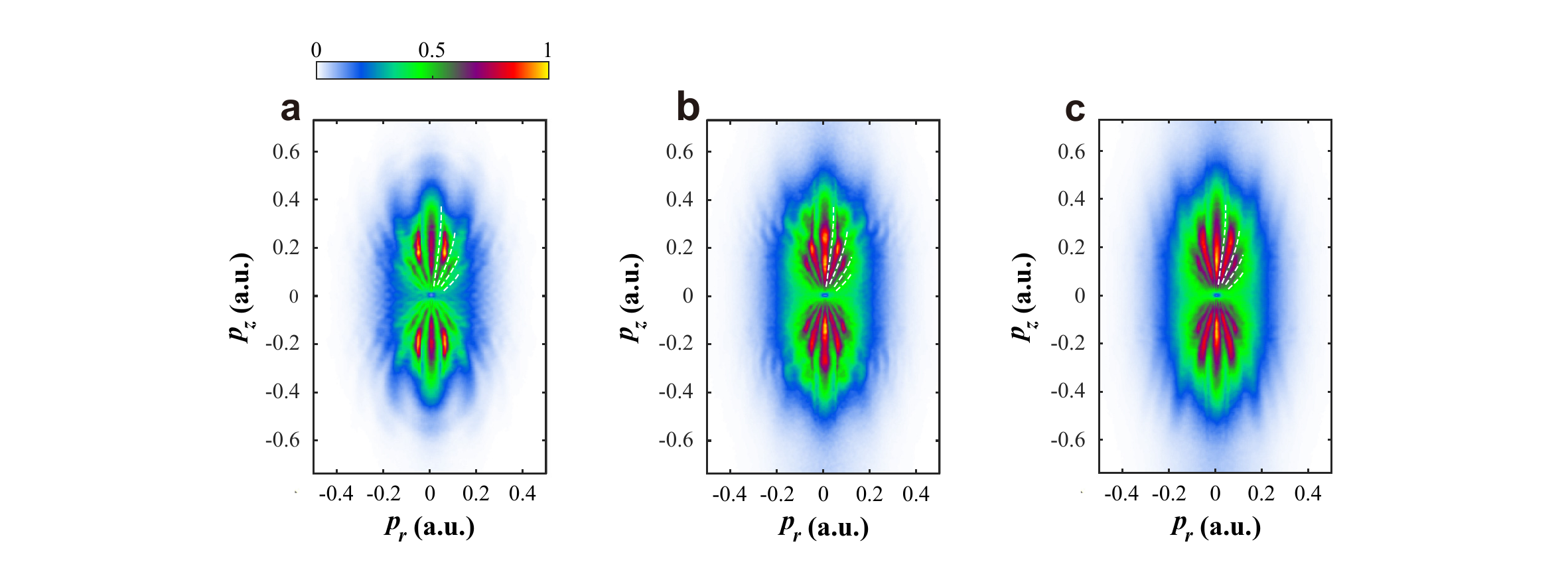}
    \caption{\textbf{Simulated PMDs using q-QTMC with BSV light} with average peak intensity of $0.53\times 10^{14}\,\mathrm{W/cm^2}$ (\textbf{a}), $0.80\times 10^{14}\,\mathrm{W/cm^2}$ (\textbf{b}) and $1.33\times 10^{14}\,\mathrm{W/cm^2}$ (\textbf{c})}
    \label{fig. 3}
\end{figure}

To explore the origin of the robustness observed in the spider-like interference pattern, we develop a quantum-light-corrected quantum trajectory Monte Carlo (q-QTMC) model (see theoretical Method for details), in which we incorporate the quantum nature of the driving field into the QTMC model \cite{2014-li-classicalquantumcorrespondenceabovethreshold}. This model successfully reproduces the experimentally observed spider-like fringes under BSV driving, as highlighted by the dashed lines in Figures \ref{fig. 3}a-c. In the simulation, we construct an ensemble of intensities $I(\alpha)$ sampled from the Husimi distribution $Q(\alpha)$, such that the ensemble-averaged pulse energy $\braket{E}=K\braket{I}=K\sum_\alpha Q(\alpha)I(\alpha)$ matches the experimental value. Under our experimental conditions, average pulse energies of $3.3,\, 5.0,\, 8.3\,\mathrm{\mu J}$ correspond to average peak intensities of $0.53,\, 0.80,\, 1.33\times 10^{14}\,\mathrm{W/cm^2}$, respectively. This model successfully reproduces the experimentally observed spider-like fringes under BSV driving, as highlighted by the dashed lines in Figures \ref{fig. 3}a-c. The q-QTMC model provides an intuitive simulation of the spider-like structure.

\begin{figure}[htbp]
    \centering
    \includegraphics[width=\textwidth]{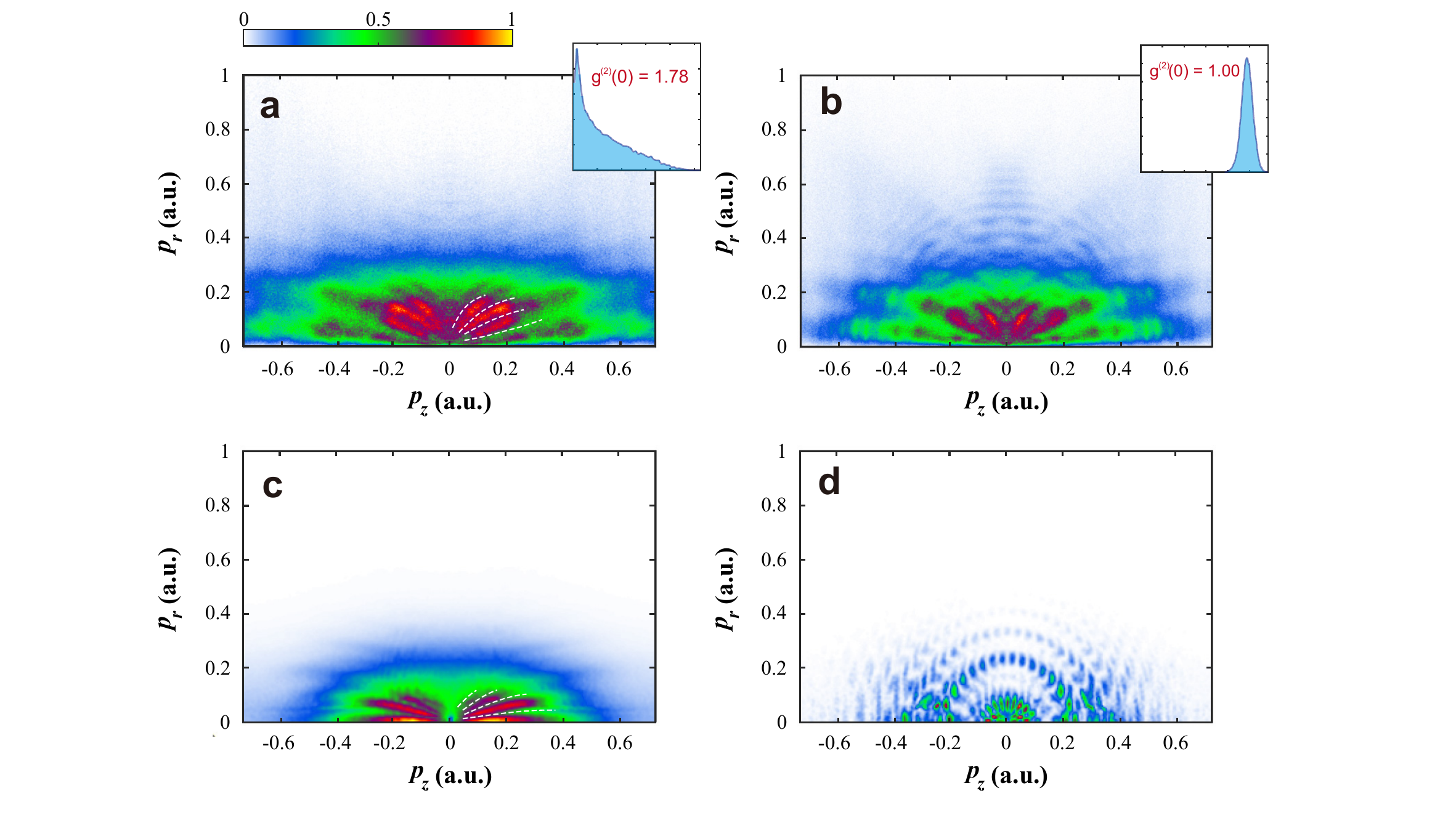}
    \caption{\textbf{Measured PMD of $p_r$ vs $p_z$ for strong-field ionization driven by BSV and coherent-state light, along with simulation result.} \textbf{a}, Experimental PMD driven by BSV with average pulse energy of $10\, \mathrm{\mu J}$ (the averaged intensity~$1.60\times 10^{14}\,\mathrm{W/cm^2}$). \textbf{b}, Experimental PMD driven by the coherent-state light at the wavelength of $1600 \,\mathrm{nm}$ with intensity of $1.0\times 10^{14}\,\mathrm{W/cm^2}$. \textbf{c}, Simulated 2D PMD using q-QTMC model with the averaged intensity $1.60\times 10^{14}\,\mathrm{W/cm^2}$. \textbf{d}, Simulated PMD using q-QTMC model for the coherent-state light with intensity of $1.0\times 10^{14}\,\mathrm{W/cm^2}$.} 
    \label{fig. 4}
\end{figure}

Experimentally, photoelectron three-dimensional momenta have been measured. In order to observe the interference pattern clearly, we can plot the PMDs of $p_r = \sqrt{p_x^2+p_y^2}$ vs $p_z$ for the experiment. In the q-QTMC model, the photoelectron momentum distributions are calculated by coherent superposition of three-dimensional trajectories onto a two-dimensional virtual detector\cite{2014-li-classicalquantumcorrespondenceabovethreshold}, which can be directly compared with the measurements. In Fig. \ref{fig. 4}a, we present the measured $p_r$ with respect to $p_z$ for the BSV light at average pulse energy up to $10\,\mathrm{\mu J}$ with $g^{(2)}(0)$=1.78. One can see that the spider structure is even more evident. The simulated PMD with the q-QTMC model (Fig. \ref{fig. 4}c) agrees better with the experimental measurement $p_r$ vs $p_z$.

For comparison, we also examine how the PMD differs under coherent-state driving fields. In Fig. \ref{fig. 4}b, we present the measured PMD driven by a $1600\,\mathrm{nm}$ coherent-state light at an intensity of $1.0\times10^{14}\,\mathrm{W/cm^2} $. As expected, the measured second-order correlation $g^{(2)}$ of coherent-state light is $1.003$ (up right corner in Fig. \ref{fig. 4}b). Obviously, a striking contrast emerges between the coherent-state and the BSV driven cases (Fig. \ref{fig. 4}a). The PMD with the coherent-state light reveals pronounced ATI rings and carpet-like interference patterns\cite{2012-korneev-interferencecarpetsabovethreshold,2014-moller-offaxislowenergystructures}. These interference features are dramatically suppressed under BSV driving. In contrast, the spider-like fringes not only survive but become even more prominent, as highlighted by dashed lines in Fig. \ref{fig. 4}a. This difference suggests that quantum fluctuations fundamentally alter the formation and visibility of specific interference structures in strong-field ionization, potentially favoring those that are less sensitive to phase variations or energy blurring.

In Fig. \ref{fig. 4}d, we show the simulated PMD with coherent-state driving at the intensity of $10^{14}\,\mathrm{W/cm^2}$ \cite{2014-li-classicalquantumcorrespondenceabovethreshold} using the QTMC model. The QTMC simulation reproduces the ATI rings and carpet-like features on PMD. In contrast, the BSV driving (Fig. \ref{fig. 4}c) leads to a selective survival of the spider-like pattern and concurrent suppression of other interference structures. This behavior reinforces the notion that such photoelectron holography exhibits intrinsic robustness under quantum-fluctuating light fields. As seen in the good agreement between q-QTMC simulations and experimental observations, we then adopt the model to capture the essential physics.

\begin{figure}[!htbp] 
    \centering
    \includegraphics[width=\textwidth]{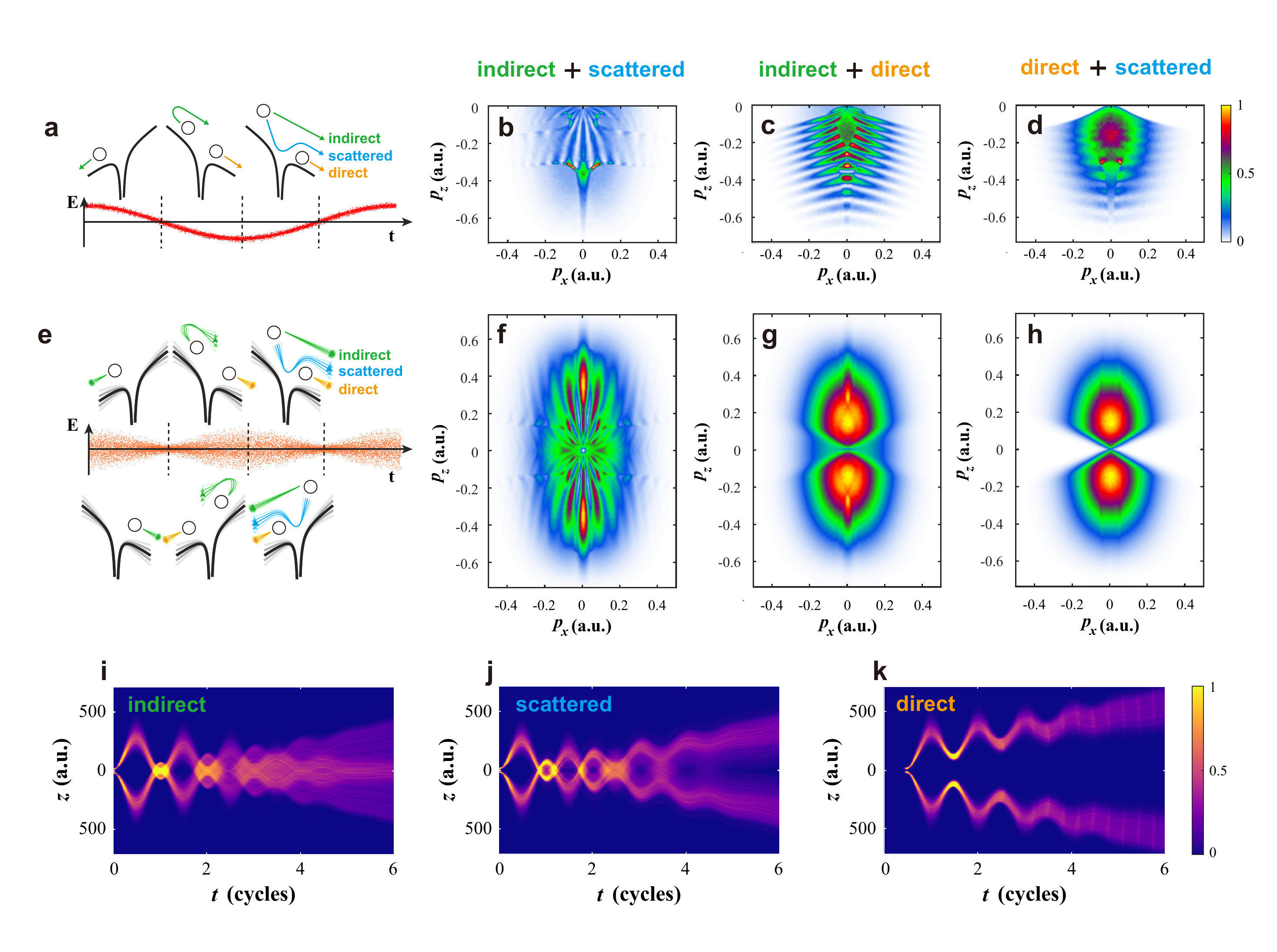}
    \caption{\textbf{Subcycle dynamics of strong-field ionization in the coherent-state and BSV light.} \textbf{a}, The ionization dynamics in an optical cycle of a coherent-state pulse. Depending on the ionization time, the released photoelectrons experience different dynamics for each quarter cycle. Electrons ionized in the first quarter-cycle follow indirect or scattered trajectories, while those ionized in the second quarter-cycle undergo direct ionization without returning to the parent ions and rescattering. These distinct electron trajectories yield similar final momenta, allowing their interference in the PMDs. \textbf{b–d}, The photoelectron interference structures with the coherent-state driving, \textbf{b}, The indirect–scattered electron interference pattern; \textbf{c}, The indirect–direct electron interference pattern; \textbf{d}, The direct–scattered interference pattern. \textbf{e}, The ionization dynamics in an optical cycle of BSV pulse. A representative BSV field reveals pronounced subcycle amplitude fluctuations. \textbf{f–h}, The photoelectron interference structures with BSV driving. Under BSV driving, quantum field fluctuations affect each interference channel differently: \textbf{f}, the indirect–scattered electron interference pattern of BSV. The spider-like structure of indirect–scattered interference is not only preserved but notably enhanced; \textbf{g}, The indirect–direct electron interference pattern of BSV; \textbf{h}, The direct–scattered electron interference pattern of BSV. \textbf{i–k}, The real-time evolution of indirect (\textbf{i}), scattered (\textbf{j}), and direct (\textbf{k}) electrons under BSV driving along the polarization axis.}
    \label{fig. 5}
\end{figure}

The motion of free and bound electrons driven by BSV light has been theoretically studied using a toy model \cite{2024-eventzur-motionchargedparticles}. Under our experimental condition, the energy of photoelectrons is mainly populated below $2U_p$ for strong-field ionization of atoms. The contribution of backward-scattered electrons to the high-energy region is negligible, and thus the interference channels related to the backward-scattered electrons can be ignored\cite{2014-li-recollisioninducedsubcycleinterference}. Using the q-QTMC framework, we can classify the contributed trajectories within an optical cycle into three categories: indirect, scattered, and direct electrons. When two types of these electron trajectories acquire the same final momentum, the interference between them will be expected. Taking full advantage of the q-QTMC model, we provide an intuitive interpretation of the enhancement of the spider-structure with BSV light. In Fig. \ref{fig. 5}a, we illustrate the ionization dynamics in an optical cycle of coherent-state light. For the classical coherent-state light, the oscillating electric field follows a sinusoidal curve with a well-defined amplitude. Within the simple-man picture of tunneling ionization, most of the electrons can be classified as direct electrons, indirect electrons, and scattered electrons depending on the ionization instant. Both indirect and scattered electrons are ionized during the first quarter cycle of the coherent-state driving field (Fig. \ref{fig. 5}a). The indirect electrons propagate away from the core with negligible interaction, while the scattered electrons revisit and scatter off the parent ion. Thus, the scattered electrons encode the structural and dynamical information about the atomic and molecular structures. Consequently, the interference between indirect electrons and scattered electrons has been termed photoelectron holography\cite{2011-huismans-timeresolvedholographyphotoelectrons}. In contrast, the direct electrons are ionized during the second quarter-cycle and are directly accelerated to the detector without returning to the parent ion. Each type of trajectory contributes differently to the formation of interference patterns. The interference between indirect and scattered trajectories gives rise to the characteristic spider-like fringes, as shown in Fig. \ref{fig. 5}b. The interference between indirect and direct trajectories leads to the time double-slit structure\cite{2006-arbo-timedoubleslitinterferences}, as shown in Fig. \ref{fig. 5}c. In Fig. \ref{fig. 5}d, the interference pattern for direct and scattered electrons gives rise to yet another interference pattern.

However, the photoelectron interference arising from these trajectories under BSV illumination differs significantly from that driven by the coherent-state light. In Fig. \ref{fig. 5}e, we illustrate the electron dynamics within a single optical cycle of BSV light. Unlike coherent-state light, BSV is a quantum state with a vanishing average electric field but with pronounced quantum fluctuations. The ionization dynamics differ remarkably from those under the coherent-state illumination. In Figures \ref{fig. 5}f-\ref{fig. 5}h, we present the interference patterns resulting from the same types of trajectory pairs: indirect and scattered electrons (f), indirect and direct electrons (g), and direct and scattered electrons (h). Accordingly, one can observe that the direct-indirect and direct-scattered interference patterns observed in the coherent-state light are substantially suppressed, whereas the spider-like fringes not only persist but are remarkably enhanced.

This coherence preservation is nontrivial, as it reveals that the selective preservation of photoelectron interference originates from intrinsic electron trajectory pairs with BSV. In reality, the surviving photoelectron hologram arises from those electrons ionized within the same quarter-cycle, whereas the interference of those electrons with different ionization times is much suppressed. This contrast suggests that the underlying mechanism is rooted in the correlation between electron trajectories, e.g., electrons released within the same subcycle are simultaneously exposed to the close quantum fluctuations, allowing them to accumulate phase in a synchronized manner. Consequently, their mutual coherence is less susceptible to quantum-induced dephasing.

To elucidate the underlying mechanism, we have simulated the representative indirect, scattered, and direct electron trajectories, as shown in Figures \ref{fig. 5}i-k. One can find that the robustness of the spider-like interference originates from the intrinsic similarity between the two contributing quantum trajectories of the indirect and scattered electrons, which are released within the same subcycle and undergo nearly synchronous evolution under the combined laser and Coulomb fields. This intrinsic similarity ensures that their phase difference remains relatively insensitive to intensity fluctuations, resulting in dynamically protected coherence. In contrast, the trajectory pairs with significantly different ionization times—such as the direct and indirect electrons—accumulate phase differences that are highly sensitive to the instantaneous field variations, rendering them more susceptible to dephasing.

Remarkably, this intrinsic robustness of photoelectron holography against strong quantum fluctuation suggests that quantum noise—typically viewed as a detrimental factor—can instead be harnessed as a resource for ultrafast imaging. Specifically, strong-field ionization inherently generates multiple interfering pathways. Various interference structures coexist and would obscure the desired holographic signal. Under quantum illumination, however, the intrinsic correlation between holographic electron trajectories selectively preserves their coherence while suppressing competing interferences. As a result, the holographic structure emerges more clearly. These findings point out a new class of coherence-protected observables in strong-field physics and highlight photoelectron holography for ultrafast imaging of molecular structure and electron dynamics under quantum-fluctuating conditions.

\section*{Conclusion}

Strong-field ionization of atoms has been experimentally studied. Our study reveals that photoelectron holography under BSV driving does not merely survive in quantum noise—it thrives within it, while other photoelectron interferences are erased.  This selective enhancement originates from a hidden order in the ionization dynamics with quantum light: the tunneling electron trajectories that share the same birth times and evolve in a correlated manner retain robust coherence, despite strong intensity fluctuations. Conversely, the interference features that lack this intrinsic synchrony succumb to dephasing and vanish from the photoelectron momentum distributions. Crucially, this is not a passive robustness. It is an active filtration, i.e., the quantum field does not indiscriminately degrade coherence, but selectively preserves it along dynamically correlated pathways. In this way, BSV acts as a quantum filter, projecting out observables that are resilient not because they are immune to noise, but because their coherence is encoded in the system’s dynamical order. Through joint experimental observations and q-QTMC modeling, we identify these surviving structures as quantum-coherence-protected features, revealing a new operational regime of strong-field physics—where coherence is not externally engineered but intrinsically sustained through the dynamic correlations between quantum light and electron motion.

Looking forward, this selective survival of intrinsically correlated electron trajectories under quantum-field drive opens a promising path for quantum-enhanced ultrafast atomic and molecular imaging. In particular, photoelectron holography driven by bright squeezed light leverages quantum-engineered strong fields to control the timing of electron emission and reduce phase noise, thereby enhancing the visibility and coherence of holographic imaging. Such an effect significantly improves the resolution and sensitivity of molecular imaging, potentially enabling quantum-enhanced strong-field spectroscopy. By tailoring the quantum properties of the driving light, i.e., squeezing strength, mode composition, and temporal correlations, it may become feasible to sculpt specific interference features that are robust to environmental noise, or even to harness quantum fluctuations. These capabilities could be transformative for strong-field and attosecond science, enabling precise probing of ultrafast electron dynamics, inner-shell motion\cite{2002-drescher-timeresolvedatomicinnershell}, or field-induced molecular processes\cite{2025-long-hydrogenmoleculardissociation} at the atomic and molecular scale. More broadly, our findings point toward a new paradigm in quantum imaging—one in which quantum fluctuations are not merely treated as noise but actively re-engineered as a resource—laying conceptual foundations for future quantum technologies.

\linenumbers
\section*{Methods}
\captionsetup[figure]{labelfont={bf},labelformat={default},labelsep=period,name={Extended Figure}}

\subsection*{Experimental methods}

\subsubsection*{BSV generation and Characterization}

\begingroup
\setcounter{figure}{0}
\renewcommand{\figurename}{Extended Figure}
    \begin{figure}[htbp]
        \centering
        \includegraphics[width=\textwidth]{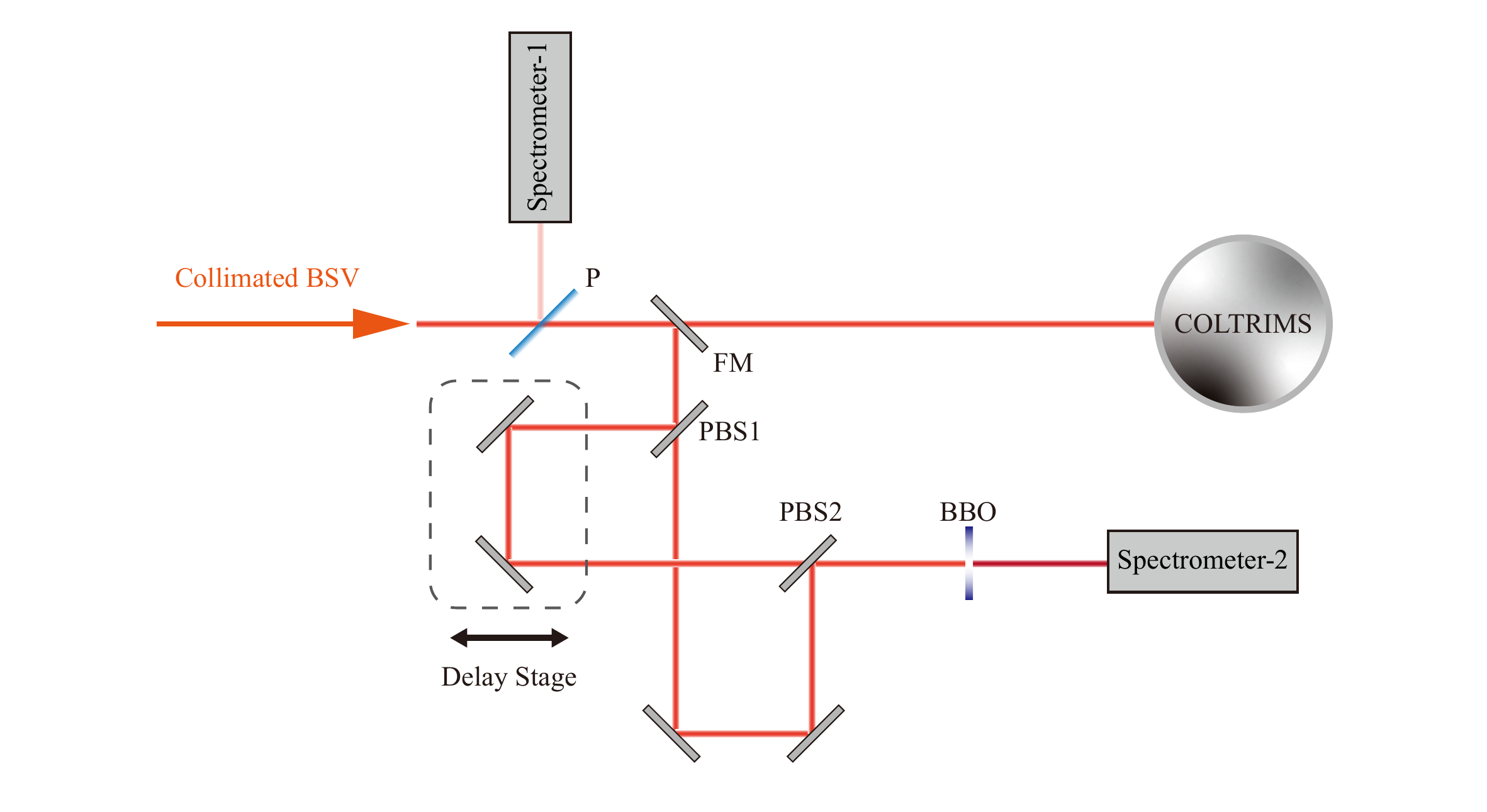}
        \caption{\textbf{Experimental setup for characterizing the BSV source.} 'P': glass plate; 'FM': flipping mirror; 'PBS': polarizing beam splitter; 'BBO': 1-mm-thick type-I beta-barium borate crystal; 'Spectrometer-1': AvaSpec-NIR256-2.5-HSC-EVO, with a spectral range of $1000\sim2500\,\mathrm{nm}$; 'Spectrometer-2': AvaSpec-ULS4096CL-EVO, with a spectral range of $200\sim1100\,\mathrm{nm}$.} 
        \label{extended fig. 1}
    \end{figure}
\endgroup

We generate the BSV pulses through type-I spontaneous parametric down-conversion (SPDC). The pump source is an amplified Ti:sapphire femtosecond laser system operating at a central wavelength of $800 \,\mathrm{nm}$, with a pulse duration of $25 \,\mathrm{fs}$ and a repetition rate of $3\, \mathrm{kHz}$. It delivers tunable single-pulse energies up to approximately $700 \,\mathrm{\mu J}$. BSV is generated via collinear, frequency-degenerate type-I SPDC in a pair of 2-mm-thick BBO crystals placed symmetrically about the common focus of an $f=1000 \,\mathrm{mm}$ lens. The crystals are angularly tuned to achieve phase matching around $1600 \,\mathrm{nm}$. A long-pass filter (cutoff at $1180 \,\mathrm{nm}$) is inserted after the crystal pair to remove residual $800\, \mathrm{nm}$ pump effectively. An $f=600\, \mathrm{mm}$ lens is used to collimate the BSV.

\newpage
\begingroup
\setcounter{figure}{1}
\renewcommand{\figurename}{Extended Figure}
    \begin{figure}[htbp]
        \centering
        \includegraphics[width=\textwidth]{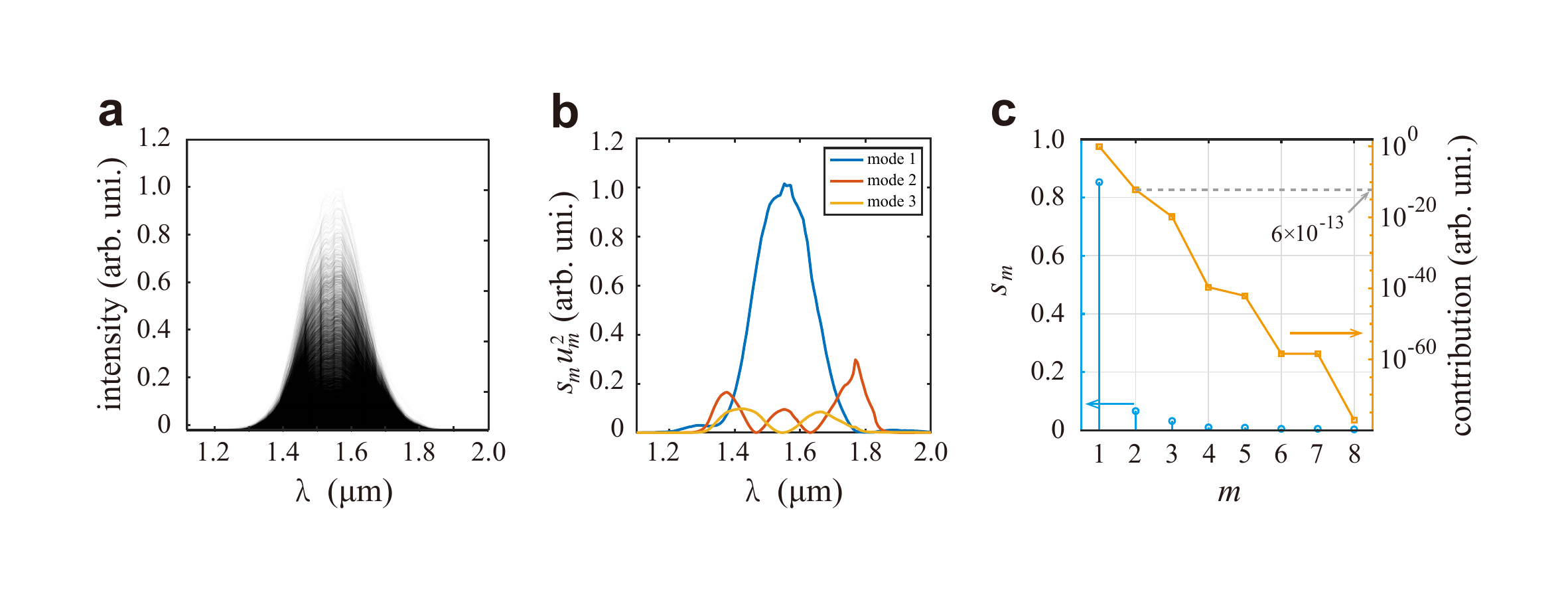}
        \caption{\textbf{Shot-by-shot measurement and modal decomposition of the BSV.}  \textbf{a}, Shot-by-shot spectra of $20,000$ BSV pulses measured with Spectrometer-1. \textbf{b}, Schmidt mode decomposition of the BSV spectra. The first mode dominates, exhibiting a single-peaked spectrum. The second and third modes, multiplied by a factor of $5$ for better visibility, display multi-peaked structures. \textbf{c}, The weights of each mode (left axis), along with their ionization contributions defined as the product of modal intensity and ADK ionization rate (right axis).} 
        \label{extended fig. 2}
    \end{figure}
\endgroup

As shown in the Extended Figure \ref{extended fig. 1}, to characterize the temporal, spectral, and statistical properties of BSV pulses, we implemented shot-by-shot measurements immediately after the generation. A low-reflectivity glass plate was used to pick off a small fraction of the BSV beam, which was sent into a spectrometer covering the $1000\sim2500\,\mathrm{nm}$ range, and spectra were recorded in the Extended Figure \ref{extended fig. 2}a. For each laser shot, the spectrometer recorded a spectrum $I(\lambda)$. Taking into account the spectrometer response function $S(\lambda)$ and the reflection coefficient $R$ of the glass plate, the photon number per pulse was obtained $N=\frac{1}{R}\int d\lambda\frac{I(\lambda)}{S(\lambda)}$. By recording spectra from $20,000$ consecutive pulses, we obtained the photon-number statistics at each BSV intensity level and computed the second-order intensity correlation function in a large-photon-number limit as $g^{(2)}=\left<N^2\right>/\left<N\right>^2$.

In parallel, we performed a spectral Schmidt mode analysis on the measured ensemble of spectra\cite{2025-barakat-simultaneousmeasurementmultimode}. Specifically, from the spectra $I(\lambda)$ we construct the wavelength–wavelength intensity-fluctuation covariance matrix $\mathrm{Cov}(\lambda,\lambda^\prime)=\left<\Delta I(\lambda)\Delta I(\lambda^\prime)\right>$, and normalize it to obtain a dimensionless correlation kernel $C(\lambda,\lambda^\prime)$. We then evaluate its matrix square root $K=\sqrt{C}$ and perform a singular value decomposition,
\begin{equation}
\begin{aligned}
\label{}
K(\lambda,\lambda^\prime)=\sum_m s_mu_m(\lambda)u_m^*(\lambda^\prime).
\end{aligned}
\end{equation}
This procedure yields an orthonormal set of spectral modes $u_m(\lambda)$ that diagonalize the experimentally accessible second-order intensity correlations, together with the corresponding modal weights $s_m$. This analysis allowed us to quantify the modal decomposition of the BSV pulse. Under our experimental conditions, the BSV was pumped in the high-gain regime. The effective number of Schmidt modes increases modestly with the pump power, leading to a gradual decrease in $g^{(2)}$\cite{2014-allevi-statisticstwinbeamstates}. Nevertheless, the strong-field ionization process is highly nonlinear in nature.  For the highest BSV intensity used in this work, its frequency Schmidt modal decomposition result is shown in the Extended Figure \ref{extended fig. 2}b. To examine the contribution of each mode in the ionization event, we can calculate the relative contribution by multiplying the mode weight and the ADK ionization rate. As shown in the Extended Figure \ref{extended fig. 2}c, the Schmidt modes beyond the first account for less than $10^{-12}$ of the total ionization events. Therefore, the contribution of photoelectron yields is mainly from the first-order Schmidt mode in our experiments, while the contributions from higher orders remain negligible.

\begingroup
\setcounter{figure}{2}
\renewcommand{\figurename}{Extended Figure}
    \begin{figure}[htbp]
        \centering
        \includegraphics[width=\textwidth]{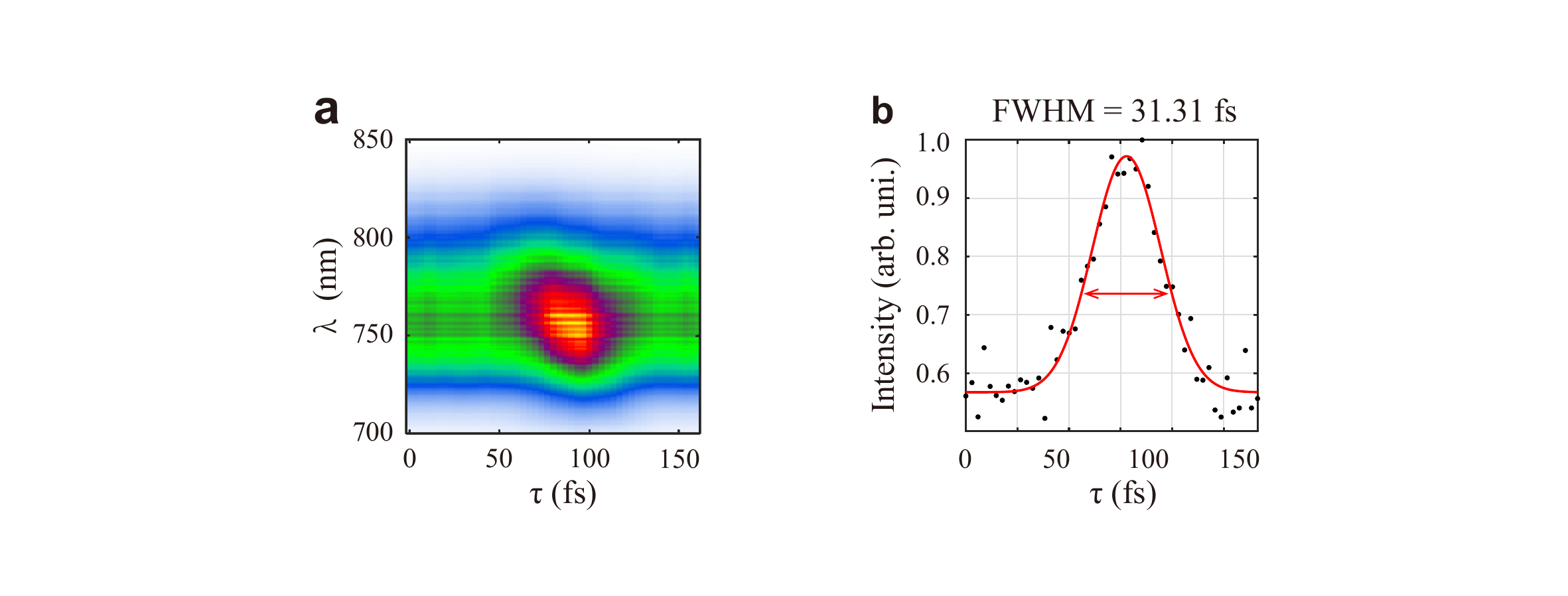}
        \caption{\textbf{Results of FROG measurement.}  \textbf{a}, Time–wavelength spectrogram obtained from FROG measurement. \textbf{b}. Temporal profile retrieved by integrating over wavelength and fitting the intensity envelope, yielding a pulse duration of $\Delta t = \mathrm{FWHM} = 31.31\,\mathrm{fs}$.} 
        \label{extended fig. 3}
    \end{figure}
\endgroup

To examine the pulse duration of the BSV, we have utilized a frequency-resolved optical gating (FROG) method. The BSV beam was redirected into a FROG setup via a flipping mirror, then split by a polarizing beamsplitter into two orthogonal components. After traveling through two arms, the beams were recombined at another polarizing beamsplitter and focused onto a BBO crystal for second-harmonic generation (SHG). By scanning the delay between the two arms and recording the SHG spectra (Extended Figure \ref{extended fig. 3}a), we reconstructed the BSV temporal envelope using a standard retrieval algorithm shown in Extended Figure \ref{extended fig. 3}b. The retrieved pulse duration at the input of the interaction region was approximately $31.31\,\mathrm{fs}$.

\subsubsection*{Strong-field ionization experiment with BSV}

PMDs of ionization with BSV are measured using a COLTRIMS system\cite{2003-ullrich-recoilionelectronmomentum,2000-dorner-coldtargetrecoil}. The BSV pulses are focused into a pulsed supersonic xenon gas jet inside a high-vacuum chamber ($5\times 10^{-11}\,\mathrm{mbar}$) by a concave mirror of focal length $75\,\mathrm{mm}$. The backing pressure is kept at $0.7\, \mathrm{atm}$ to maintain sufficient ionization yield while avoiding space-charge effects. The electrons are extracted and guided by a well-calibrated combination of electric and magnetic fields toward a microchannel plate (MCP) detector equipped with a delay-line detector (DLD). This setup allows for time- and position-resolved detection of each electron, thereby enabling full reconstruction of the three-dimensional momentum vector on a single-event basis. By accumulating millions of events, we obtain statistically significant three-dimensional PMDs for each experimental condition.

For the experiment with coherent-state light, the BSV beam is replaced by coherent-state mid-IR pulses at $1600\,\mathrm{nm}$ wavelength generated from an optical parametric amplifier (OPA) pumped by the same laser system. The OPA output is passed through the same optical path and focused onto the xenon target under identical experimental geometry.

\subsection*{Theoretical method: quantum-light-corrected quantum trajectory Monte Carlo (q-QTMC) theory}

To understand the photoelectron momentum distributions in the presence of BSV, we have developed a quantum-light-corrected quantum trajectory Monte Carlo (q-QTMC) model. Briefly, for strong-field ionization with the coherent-state field, each electron trajectory is weighted by the Ammosov-Delone-Krainov (ADK) theory \cite{1986-ammosov-tunnelionizationcomplex,1991-delone-energyangularelectron}, $W(t_0,v_\perp^j)= W_0(E(t_0))\times W_1(v_\perp^j)$\cite{2014-li-classicalquantumcorrespondenceabovethreshold}, in which $W_0(t)=|(2I_p)^2/E(t)|^{2/\sqrt{2I_p}-1}\exp{-2(2{I}_{p})^{3/2}/|3 {E}({t})|}$ determines the ionization rate with respect to the ionization instant, and $W_1(v_\perp^j)=\sqrt{2I_p}/E(t)\exp{-\sqrt{2I_{p}}(v_{\perp}^{j})^2/\left|{E}({t})\right|}$ determines the initial transverse momentum distribution, where $E(t)$ is the instantaneous strength of the laser field and $I_p$ is the ionization potential. The electron tunneling exit is along the instantaneous direction of the light field, and its value is given by $I_p/|{E}(t)|$. After sampling all the electrons, their classical motion outside the barrier is governed by the Newtonian equation $\ddot{\textbf{r}} = - {E}(t)- \textbf{r}/r^3$ until the laser is turned off, where $r$ is the distance from the electron to the nucleus. To solve the Newtonian equations more precisely near the nucleus, we use an explicit method that automatically selects the solution algorithm when solving ordinary differential equations. To interpret the photoelectron holography,  the trajectory phase is given by a Feynman path-integral approach $\Phi_j=\int^\infty_{t_0}dt\left[\frac{\textbf{p}(t)^2}{2}+I_p-V\left(\textbf{r}(t)\right)\right]$, where $p(t)$ is the electron's instantaneous velocity, $I_p$ is the ionization potential, and $V(\textbf{r}(t))$ is the atomic Coulomb potential. The final PMD is then obtained by coherently summing over all trajectories that reach the same final momentum $M(\textbf{p})=\left|\sum_j \sqrt{W(t_0,v_\perp^j)}\exp{-i\Phi_j}\right|^2$. 

To extend the QTMC framework to quantum light, we incorporate the quantum statistical nature of the bright squeezed vacuum (BSV) field via the reduced electronic density matrix formalism in the joint system of light and atom. In the framework, we have considered several assumptions: i) negligible atom–field entanglement; ii) small single-photon amplitude limit, and iii) negligible fluctuations of the coherent-state basis in the high-photon-number regime. The total atom–field density matrix is thus expressed as an incoherent mixture over the coherent-state projections\cite{2023-gorlach-highharmonicgenerationdriven,2023-eventzur-photonstatisticsforceultrafast}. By tracing out the field degrees of freedom, the reduced electronic density matrix is given by,
\begin{equation}
\begin{aligned}
\label{eq2-8}
\rho_e(t)
=\int d^2\alpha Q\left(\alpha\right)\ket{\phi_\alpha(t)}\bra{\phi_\alpha(t)}.
\end{aligned}
\end{equation}
where $Q(\alpha)$ is the Husimi $Q$-function describing the quantum state of the light, and $\ket{\phi_\alpha(t)}$ is the electronic wavefunction evolved under a classical field corresponding to the coherent amplitude $\alpha$. The final PMD under quantum-light driving is obtained by projecting $\rho_e$ onto the ionized state via a projection operator $\Pi$, yielding:
\begin{equation}
\begin{aligned}
M_q(\textbf{p}) &= \bra{\textbf{p}} \Pi \rho_e(t\to\infty) \Pi \ket{\textbf{p}} 
\\&= \int d^2\alpha \, Q(\alpha) M_\alpha(\textbf{p}),
\end{aligned}
\end{equation}
where $M_\alpha(\textbf{p})$ denotes the PMD resulting from a coherent-state field realization of amplitude $\alpha$.

\section*{Acknowledgments}

This work is supported by the National Key R$\&$D Program (No. 2022YFA1604301) and Natural Science Foundation of China (Nos. 12334013 and 92250306).

\section*{Author Contributions}

H.L., X.L., and P.L. designed and performed the experiments. H.L. built the theoretical model and conducted simulations. H.L. and Y.L. analyzed the data. H.L. and Y.L. drafted the paper with extensive input from all authors. Y.L. conceived the idea and supervised the project.

\section*{Additional information}

\subsection*{Data availability} The datasets generated during and/or analysed during the current study are available from the corresponding author on reasonable request.

\subsection*{Code availability} All codes used for the analysis and production of results of the current study are available from the corresponding author on reasonable request.

\end{document}